\documentclass[preprint,noshowpacs,preprintnumbers,amsmath,amssymb,prl,superscriptaddress]{revtex4-2}

\usepackage{graphicx}
\usepackage{dcolumn}
\usepackage{bm}
\usepackage{epsfig}
\usepackage{natbib}
\usepackage{color}
\usepackage{xcolor}
\usepackage{lineno}
\usepackage{lipsum}

\newcommand{\VBG}{V_\text{BG}} 
\newcommand{\Rxx}{R_\text{xx}} 
\newcommand{\Tc}{T_\text{c}} 

\begin{document}
\title{Re-entrant superconductivity at an oxide heterointerface}
\today
\author{D.~Maryenko}
\email{maryenko@riken.jp}
\affiliation{RIKEN Center for Emergent Matter Science (CEMS), Wako 351-0198, Japan}
\author{M.~Kawamura}
\affiliation{RIKEN Center for Emergent Matter Science (CEMS), Wako 351-0198, Japan}
\author{I. V. Maznichenko}
\affiliation{Institute of Physics, Martin Luther University Halle-Wittenberg, 06120 Halle, Germany}
\author{S. Ostanin} 
\affiliation{Institute of Physics, Martin Luther University Halle-Wittenberg, 06120 Halle, Germany}
\author{D. Zhang}
\affiliation{RIKEN Center for Emergent Matter Science (CEMS), Wako 351-0198, Japan}
\affiliation{State Key Laboratory of Low Dimensional Quantum Physics and Department of Physics, Tsinghua University, Beijing, China}
\author{M.~Kriener}
\affiliation{RIKEN Center for Emergent Matter Science (CEMS), Wako 351-0198, Japan}
\author{V.~K.~Dugaev}
\affiliation{Department of Physics and Medical Engineering, Rzesz\'{o}w University of Technology, 35-959 Rzesz\'{o}w, Poland}
\author{E.~Ya.~Sherman}
\affiliation{Department of Physical Chemistry, University of the Basque Country UPV/EHU, Apartado 644, Bilbao 48080,  Spain}
\affiliation{Ikerbasque, Basque Foundation for Science, Bilbao, Spain}
\author{A.~Ernst}
\affiliation{Institute for Theoretical Physics, Johannes Kepler University, 4040 Linz, Austria}
\affiliation{Max Planck Institute of Microstructure Physics, D-06120 Halle, Germany}
\author{M.~Kawasaki}
\affiliation{RIKEN Center for Emergent Matter Science (CEMS), Wako 351-0198, Japan}
\affiliation{Department of Applied Physics and Quantum-Phase Electronics Center (QPEC), The University of Tokyo, Tokyo 113-8656, Japan}

\maketitle


\textbf{A magnetic field typically suppresses superconductivity by either breaking Cooper pairs via the Zeeman effect or inducing vortex formation. However, under certain circumstances, a magnetic field can stabilize superconductivity instead \cite{FieldInducedSC1984,FieldInducedSC2004}. This seemingly counterintuitive phenomenon is associated with magnetic interactions \cite{ExchangeField1962} and has been extensively studied in three-dimensional materials \cite{URhGe2005, UTe2_2019, EuFeCoAs, CeRhAs, OrganicSC}. By contrast, this phenomenon, hinting at unconventional superconductivity, remains largely unexplored in two-dimensional systems, with moiré-patterned graphene being the only known example  \cite{MoirePatternRSC}. Here, we report the observation of re-entrant superconductivity (RSC) at the epitaxial (110) - oriented LaTiO$_3$-KTaO$_3$ interface. This phenomenon occurs across a wide range of charge carrier densities, which, unlike in three-dimensional materials, can be tuned \textit{in-situ} via electrostatic gating. We attribute the re-entrant superconductivity to the interplay between a strong spin-orbit coupling and a magnetic-field driven modification of the Fermi surface. Our findings offer new insights into re-entrant superconductivity and establish a robust platform  for exploring novel effects in two-dimensional superconductors. }
 
The behavior of superconductors under magnetic fields offers profound insights into the properties of the superconducting phase and related emergent quantum states. Magnetic fields conventionally suppress superconductivity by breaking the coherence of Cooper pairs. However, in rare cases, they can stabilize the superconducting state. One notable example is the Jaccarino-Peter effect~\cite{ExchangeField1962}. In essence, an applied external magnetic field compensates for an effective internal exchange field, which cancels the polarization of the conduction electrons.  If the attractive interaction between electrons is sufficiently strong, the superconducting phase can be stabilized in the compensation region. Originally proposed and later demonstrated in a number  of ferromagnetic materials, the RSC has been also observed in organic superconductors \cite{OrganicSC} and is discussed  in the context of the heavy fermion system UTe$_2$~\cite{Helm2024}. A different mechanism for RSC at high magnetic fields arises in heavy-fermion compounds. Here, the transition between low-field and high-field superconducting phases is accompanied by strong ferromagnetic fluctuations, which originate from the quantum instability of the ferromagnetic phase. The re-entrant superconducting phase is considered unconventional with suggested spin-triplet pairing.  
Thus, the RSC arises from the intricate interplay between multiple degrees of freedom—particularly magnetism—and has long been a paradigm studied in three-dimensional materials.
\begin{figure*}[!thb]
\includegraphics{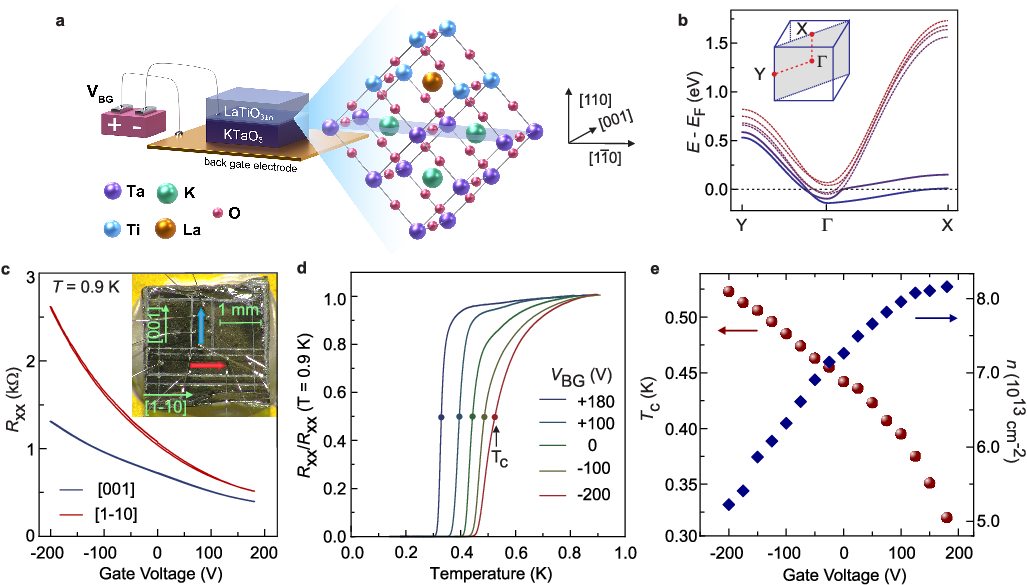}
\caption{\label{Fig1} \textbf{Properties of the (110)-oriented LaTiO$_3$-KTaO$_3$ interface}. \textbf{a)} Sketch of the interface between LaTiO$_3$ and KTaO$_3$. The backside of KTaO$_3$ acts as an electrode to apply an electrostatic field to tune the in-situ interface electronic properties. \textbf{b)} Relativistic band structure of a 2D electron gas emerging at the (110) interface,  which was simulated from first principles. Nearly flat-band regions around $E_F$ = 0  along $\Gamma$–X near the Brillouin zone boundary are formed almost  equally by highly localized $d_{z^2}$ orbitals of Ta and Ti. Solid lines are bands that are important for the theory model presented below. Dashed lines are other bands. The inset shows the unit cell in real space. \textbf{c)} The backgate voltage can effectively tune the sample resistance. Inset depicts the photograph of the device under study. The device allows to flow the current  along the [001] and [1-10] crystallographic directions independently. \textbf{d)} Superconducting transition at different backgate voltages $\VBG$. Note that the resistance is normalized to its  value. The superconducting  transition temperature $\Tc$ is defined as a temperature at which the resistance has dropped by 50$\%$ from its value at normal conducting state. \textbf{e)} Gate dependence of the superconducting transition temperature $\Tc$ and the charge carrier density $n$. } 
\end{figure*}

By contrast, the occurrence of RSC in two-dimensional systems remains largely unexplored. Two-dimensional (2D) systems provide a fundamentally distinct platform, where reduced dimensionality and broken inversion symmetry give rise to electronic states different from those in the bulk. In such systems, strong spin-orbit coupling (SOC) profoundly influences the superconducting phase. For instance, in the presence of a Rashba-type SOC, the superconducting order parameter can develop a mixed-parity state, combining both even- and odd-parity components~\cite{Rashba2001}. Despite growing interest in 2D superconducting materials, which have already revealed fascinating superconducting properties~\cite{Ye2015, NbSe2_2016, Mos2_2016, Smidman_2017, Iwasa2017,GrapheneSC2018, Ye2023},  the RSC has been observed so far only in magic angle twisted trilayer graphene~\cite{MoirePatternRSC}.  However, this system lacks strong SOC and pronounced magnetism,  leaving the origin of the re-entrant superconductivity in graphene uncertain.

Here, we report the observation of re-entrant superconductivity at the (110)-oriented epitaxial interface of LaTiO$_3$-KTaO$_3$, a system fundamentally distinct from previously studied re-entrant superconductors. Unlike other materials where RSC has been observed so far, the LaTiO$_3$-KTaO$_3$ interface does not provide clear experimental evidence for electronic correlations or intrinsic magnetism. Instead, our system features a reduced dimensionality and strong spin-orbit coupling. 
KTaO$_3$ is a cubic perovskite oxide with a conduction band derived from the 5$d$-orbitals of the Ta atoms, known for their strong spin-orbit interaction. Its interface can exhibit a sizable  Rashba- or/and Dresselhaus-type spin-orbit coupling, and its superconducting properties strongly depend on the crystal plane orientation~\cite{EuOKTOScience, EuOKTONatComm, LAOKTO}.  Recent studies, primarily on (111)-oriented interfaces, suggest the formation of unconventional superconducting phases, including evidence of a stripe phase \cite{EuOKTOScience}, spontaneous symmetry breaking~\cite{Zhang2023}, and the superconducting transition temperature $\Tc$ dependence on the in-plane direction~\cite{Hua2024}. 
\\
\\
\textbf{LaTiO$_3$-KTaO$_3$ heterostructure} Figure 1a schematically illustrates the (110)-oriented LaTiO$_3$-KTaO$_3$ interface, which is grown using pulsed laser deposition technique ~\cite{LTO-KTO2023}. In this orientation, not only the interface breaks the inversion symmetry, but also the in-plane  [001] and [1$\overline{1}$0] crystal directions are no longer equivalent. This asymmetry is reflected in the band dispersions plotted in Fig.~1b along $\Gamma$–X  and $\Gamma$–Y. For the reciprocal direction $\Gamma$-Y, which is related  to the lattice vector [1$\overline{1}$0],  all bands crossing the Fermi energy $E_F$ possess relatively high  velocities. However, for $\Gamma$-X, the two lowest conduction bands become nearly flat near the Brillouin zone boundary. This is a typical feature of the band asymmetry computed within the (110) geometry that inevitably appears as a single flat $\Gamma$–X band in insulating KTaO$_3$(110) above its band gap.  In the case of the (110) LaTiO$_3$-KTaO$_3$ interface, the two flat bands, which are formed almost equally by the $d$-states of the Ta-Ti pairs, appear around $E_F$ along $\Gamma$-X. As a result, the Fermi surface of LaTiO$_3$-KTaO$_3$ (110) is elongated along this direction, which can change the transport properties measured along [110] and [100]. The effects of the spin-orbit coupling, magnetic exchange interactions, and spin noncollinearity, which were included in the \textit{ab-initio} calculations, determine the flat-band splitting and detailed Fermi surface, as well as its spin texture (see Fig. S1 in SI).

\begin{figure*}[!thb]
\includegraphics{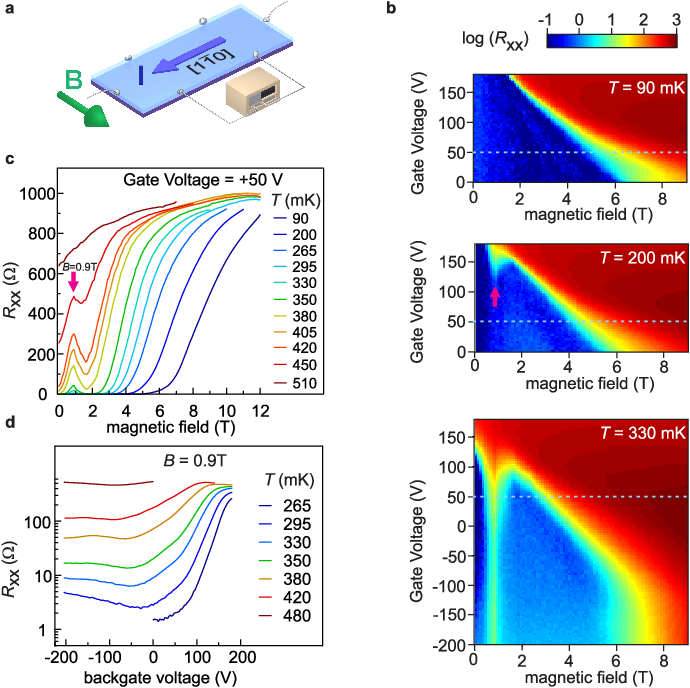}
\caption{\label{Fig2} \textbf{Experimental visualization of re-entrant superconducting state}. \textbf{a)} Scheme of the experimental arrangement. \textbf{b)} Color rendition plot of the longitudinal resistance $\Rxx$ as functions of the gate voltage and the magnetic field. The re-entrant superconductivity visualizes as the temperature increases. \textbf{c)} Magnetoresistance $\Rxx$-traces for various temperature at $\VBG=+50$ V. A resistive peak emerges at $B=0.9$~T as the temperature increases. \textbf{d)} Gate voltage dependence of the amplitude of the resistive peak at several temperatures.}  
\end{figure*}

The electronic characteristics of the interface are probed through electrical transport measurements. A photograph of the device under study is shown in the inset of panel \textbf{c}.  The current can be injected along two orthogonal in-plane crystallographic directions using independently defined conductive channels. A backgate voltage $\VBG$, applied between the backside of the KTaO$_3$ substrate and the interface, tunes the channel conductance. 
Within the applied gate voltage range of $-200$~V to $+180$~V, no measurable gate leakage current is detected. Furthermore,  the hysteresis for $\Rxx$ in gate sweeps is barely visible in panel \textbf{c}, ensuring a reliable device operation  as  a field-effect transistor. For all applied gate voltages the interface undergoes a superconducting transition. Figure 1d exemplifies the temperature dependence of $\Rxx$ near the superconducting transition for selected $\VBG$ values. For clarity of the presentation,  $\Rxx$ is normalised to its value at 0.9~K, where the superconductivity is suppressed. Panel \textbf{e} shows that the superconducting transition temperature $\Tc$ increases as $\VBG$ decreases, while the charge carrier density, estimated from the Hall effect at 0.9~K, decreases accordingly.  The gate tunability of $\Tc$ follows a similar trend to that observed in (111)-oriented KTaO$_3$ interfaces~\cite{Liu2023}. However, unlike the (111) case, we also observe a significant tunability of the charge carrier density with $\VBG$.

\begin{figure*}[!thb]
\includegraphics{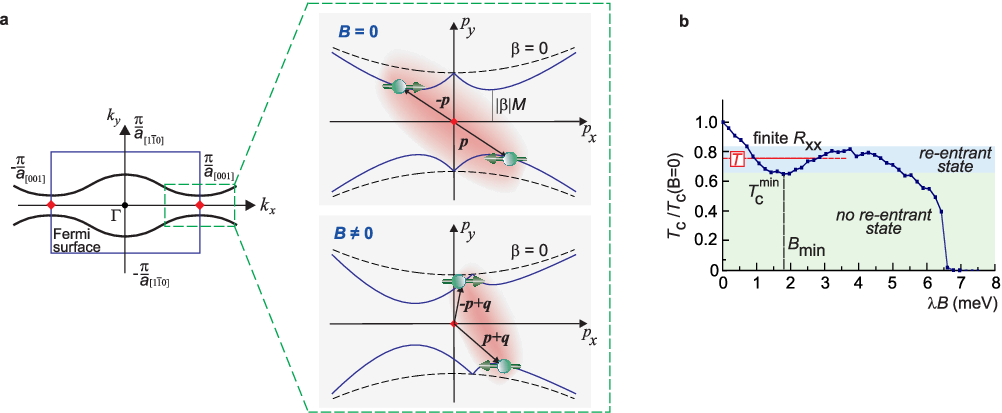}
\caption{\label{Fig4} \textbf{Theoretical model to explain the observed re-entrant superconductivity}. \textbf{a)} Left panel: Brillouin zone and Fermi surface demonstrating the van Hove singularity in the absence of spin-orbit coupling at the Brillouin zone boundary. The lattice constants are $a_{[1\overline{1}0]}$ and $a_{[100]},$ respectively. Right panel: zoom in on the states near the van Hove singularity taking into account spin-orbit coupling and magnetic field. Dashed lines correspond to the Fermi surface without spin-orbit coupling. Solid lines, marked with the corresponding spin orientations, show the deformation of the Fermi surface by spin-orbit coupling and magnetic field. The Fermi surface states contributing to the Cooper pairing are shown as filled circles. \textbf{b)}  Numerical result demonstrating the existence of the required for the re-entrant superconductivity minimum in the $T_{c}(B)-$dependence.  $T_{c}$ is normalized to its value at $B=0$.}
\end{figure*}

\textbf{Re-entrant superconductivity} The central result of this work is summarized in Figure 2, where the magnetic field is applied in-plane as sketched in panel \textbf{a}. 
The longitudinal resistance $\Rxx$ is measured at various temperatures upon sweeping the back-gate voltage at discrete magnetic field values. This approach eliminates temperature variations that could arise from sweeping the field.  Figure 2b presents a color rendition of $\Rxx$ on a logarithmic scale at 90~mK, plotted as functions of the magnetic field and the gate voltage. The critical field required to suppress superconductivity increases as $\VBG$ decreases, consistent with the dependence of $\Tc$ on the gate voltage (Fig. 1e). Notably, at this low temperatur the critical field already exceeds the laboratory magnetic field of 9~T, despite the limited gate range of  0~V and +180~V. As expected, increasing the temperature reduces the critical field at which superconductivity is destroyed. However, beyond this anticipated trend, Figure 2b reveals the emergence of a resistance cusp at $B = 0.9$~T, as indicated by the red arrow. This feature becomes more pronounced with increasing temperature, as shown in Figure 2b at $T = 330$~mK. The cusp appears consistently at $B=0.9$ T, independent of both temperature and gate voltage. Thus, upon sweeping the magnetic field, the coherent state is first suppressed (indicated by $\Rxx\neq 0$) and then re-emerges. This constitutes the phenomenon of re-entrant superconductivity. To highlight this behavior,  Figure 2c shows $\Rxx$ traces at $\VBG = +50$~V as a function of magnetic field  for various temperatures. At low temperatures, the 0.9 T resistance peak is not observable, but as the temperature increases, it gradually emerges. We have qualitatively reproduced this result in another independently grown  LaTiO$_3$-KTaO$_3$ (110)-oriented heterostructure, confirming the robustness of our observation (see Supplementary Information Sec. III).

One of the primary mechanisms that can suppress superconductivity is Zeeman splitting, which breaks Cooper pairs by aligning electron spins.  In conventional Bardeen–Cooper–Schrieffer superconductors, assuming a $g$-factor of $g=2$, the critical field is given by the Clogston–Chandrasekhar paramagnetic limit $B_{P}=1.86\times T_{c},$ where $\Tc$ is the superconducting temperature in Kelvin, and $B_{P}$ is given in Tesla \cite{Clogston, Chandrasekhar}. In our system, this places $B_{P}$ in the range from 0.6~T to 0.9~T, depending on the gate voltage.  The resistive peak at 0.9~T appears close to this paramagnetic limit. However,  unlike $T_c$, the peak position is independent of the gate voltage and its amplitude exhibits a non-monotonic dependence on $\VBG$.   As shown in Figure 2d, the peak amplitude reaches a minimum around $\VBG=-50$~V and increases again for more negative $\VBG$. This suggests that the peak is unlikely to be directly related to the Zeeman effect. Another relevant energy scale is associated with vortex formation due to the applied in-plane magnetic field. Assuming that the resistance peak at $B = 0.9$~T corresponds to the onset of flux penetration, we apply the condition $B_\parallel \xi w \cong \phi_0$, where $\phi_0$ is the magnetic flux quantum and $\xi$ is the coherence length extracted from perpendicular field measurements (Supplementary Information, Section II), to estimate the effective spatial extent $w$ of the wavefunction across the interface. This yields $w \approx 45$~nm, which significantly exceeds the typical conducting layer thickness at oxide interfaces, estimated to be around or even less than 10~nm. This inconsistency indicates that the vortex penetration is unlikely to be the origin of the resistance peak at 0.9~T. 

As described above, the emergence of superconductivity in magnetic fields has been linked to the Jaccarino-Peter effect and strong ferrromagnetic fluctuations. However, our transport experiments show no signatures of magnetic ordering. Specifically, we neither observe a hysteresis in the Hall effect nor in the longitudinal resistance, which would typically indicate the presence of effective magnetic interactions. While we cannot entirely exclude the presence of local magnetic moments near the interface - possibly arising from the oxygen vacancies \cite{LAOSTOMagnetism1, LAOSTOMagnetism2} - their effect is unlikely to be strong enough to fully explain the observed RSC phenomenon.  
A notable feature of our system is its high parallel critical field, which exceeds the perpendicular critical field by more than the order of magnitude. This suggests a crucial role of spin-orbit coupling, which can significantly enhance the critical field for Cooper pair breaking, as demonstrated in transition metal dichalcogenides \cite{Ye2015, NbSe2_2016, Mos2_2016}. 
In dichalcogenides, Ising type spin-orbit coupling locks electron spins out-of-plane and prevents their alignment with an external magnetic field.  By contrast, at the LaTiO$_3$-KTaO$_3$ interface, we expect Rashba- or Dresselhaus-type spin-orbit coupling, where the spin orientation is locked in-plane. This has the important consequence that a superconducting state with a mixture of singlet-triplet character can already be realized in zero magnetic field \cite{Rashba2001}. An applied magnetic field can modify the relative weight of singlet and triplet components, but this alone does not  explain the RSC observation in our system. 

\textbf{Discussion} The effect of re-entrant superconductivity can be qualitatively understood as a non-monotonic dependence of the superconducting transition temperature, $T_c(B)$, on magnetic field. Specifically, $T_c(B)$ exhibits a minimum, $T_c^{\rm min}$, at a certain field value $B_{\rm min}$. When the experimental temperature exceeds this minimum ($T > T_c^{\rm min}$), a finite resistive region emerges between two superconducting phases as a function of $B$. At lower temperatures ($T < T_c^{\rm min}$), this resistive interval is absent. 

Such a minimum can occur as a joint effect of the van Hove singularity obtained in \textit{ab initio} calculations (see Fig.~1b) at the Brillouin zone boundary and the Dresselhaus-like spin-orbit coupling that can be attributed to the asymmetry of the interatomic bonds at the interface electron states \cite{Krebs1996,Roessler2002, Golub2004} with a large contribution of relatively heavy Ta atoms defining the magnitude of this coupling. According to the \textit{ab initio} results, we consider electrons near the extended saddle point singularity (see Fig. 3a) with the effective mass Hamiltonian (we use $\hbar\equiv 1$ units)
\begin{equation}
H=-\frac{p_{x}^{2}}{2M}+\frac{p_{y}^{2}}{2m}+\left(\beta p_{x}+\lambda B\right) \sigma_{x},
\end{equation}
with momentum components $p_{x}$ and $p_{y}$ being centered at the zone boundary $\pi/a_{[001]}-$point (see in Fig. 3a) such that $p_{x}=k_{x}-\pi/a_{[001]},p_{y}=k_{y},$ and $M\gg m$. To demonstrate qualitatively the origin of the effect, we consider an extremely anisotropic spin-orbit coupling with $\beta$ being the coupling constant (we assume $\beta<0$, cf. Supplemental Material), $B$ the applied in-plane magnetic field, and $\lambda$ the effective system-dependent Land\'e factor. The spin-orbit coupling and magnetic field lead to the momentum-dependent spin splitting of the electron spectrum resulting in a deformation of the Fermi surface, especially pronounced in the vicinity of the van Hove singularity. We fix the chemical potential $\mu$ and the shape of the Fermi surface by choosing the upper spin-related branch with the higher energy due to spin orbit coupling as 
\begin{equation}
\frac{p_{y}^{2}}{2m}=\mu+\frac{p_{x}^{2}}{2M}-\left|\beta p_{x}+\lambda B\right|
\end{equation}
with the $\beta-$ and $B-$dependent velocity components 
\begin{eqnarray}
v_{x}&=&-\frac{p_{x}}{M}+\beta\langle\sigma_{x}\rangle, \\
v_{y}&=&\frac{p_{y}}{m}=\sqrt{2m\left(\mu-|\beta p_{x}+\lambda B|\right)+\frac{m}{M}p_{x}^2}. \nonumber    
\end{eqnarray}
In the absence of a magnetic field, the expectation value of spin is determined by spin-orbit coupling as $\langle \sigma_{x}\rangle=-{\rm sgn}\left(p_{x}\right).$ Here $p_{y}$ at the Fermi surface has two symmetric minima at $p_{x}=\pm \beta M,$ with the shift from the Brillouin zone boundary enhanced by the large mass $M.$ The corresponding Cooper pairing at $B=0$ with momenta $\mathbf{p}$ and $-\mathbf{p}$ (equivalent to the $\mathbf{k}$ and $-\mathbf{k}$ pairing in the first Brillouin zone) is shown in Fig. 3a. Nonzero magnetic field making the Fermi surface $\mathbf{p}\leftrightarrow-\mathbf{p}$ asymmetric conserves positions of the minima at $p_{x}=\pm \beta M,$ where $p_{y}^{2}/2m=\mu-M\beta^{2}/2\pm \lambda B$ and $\langle \sigma_{x}\rangle=-{\rm sgn}\left(\beta p_{x}+\lambda B\right),$ as shown in Fig. 3a. 
Notice that the minimum at $p_{x}=-\beta M>0$ exists only at a sufficiently weak magnetic field with $M\beta^{2}>\lambda B$ and disappears at stronger fields. Thus, for the upper branch of the spin-dependent spectrum the magnetic field shifts the minimum at $p_{x}=\beta M$ closer to the $x-$axis and, thus, increases the density of the states. In addition, depending on the direction of the magnetic field, its effect decreases the number of electrons with given spin direction. The asymmetry and the shift of the Fermi surface result in pairing with a nonzero total momentum dependent on the applied magnetic field. The pairing pattern shown in Fig.~3b can lead to a decrease in the transition temperature since the pairing becomes less effective.  This effect competes with the increase in the density of states at the Fermi level and leads to the minimum in the transition temperature as a function of the magnetic field. As a result, we obtain the required minimum in the $T_{c}(B)-$dependence, as shown in Fig.3b while at experimental $T<T_{c}^{\rm min}$ the interface remains superconducting up to a single critical magnetic field. This is in agreement with the experimental data.

We notice that this approach is based on the $B-$dependent shape of the Fermi surface as it can occur in the presence of a van Hove singularity and the spin-orbit coupling. It does not explicitly involve the effects of magnetic impurities except their possible impact on the $\lambda$ factor. This effect can be accompanied by other additional features related to the properties of the LaTiO$_{3}$-KTaO$_{3}$ interface. We mention here two possibilities. The momentum-dependent pairing can lead to a modification of the coupling constant and, thus, to a decrease in the transition temperature. In addition, if the superconductivity occurs due to ferromagnetic or antiferromagnetic fluctuations, possibly enhanced by a high density of states near the van Hove singularity, a magnetic field can suppress them and lead to a decrease in the transition temperature, which can be restored by the increase in the density of states and, thus, produce the re-entrant superconductivity. 


\textbf{Conclusions and outlook} 
 We have uncovered a striking re-entrant superconducting phase at the LaTiO$_3$–KTaO$_3$ interface, where the superconducting state is initially suppressed but unexpectedly re-emerges at increasing in-plane magnetic fields. This behavior challenges standard paradigms based on magnetic pair-breaking mechanisms and instead points to a novel interplay between spin-orbit coupling and a van Hove singularity near the Fermi level. Moreover, the emergence of a  low resistive peak at intermediate magnetic fields — well below the normal state resistance — suggests the presence of a transient state that separates two superconducting phases. Understanding the microscopic nature of this resistive feature may provide critical insight into superconductivity of low-dimensional systems.

Our results open a new route for exploring unconventional superconducting and finite resistivity states in engineered oxide heterostructures, where interfacial symmetry, spin-orbit interaction, and band structure can be finely tuned.  The gate-tunability of our system provides a powerful platform to probe the microscopic origin of re-entrant behavior and phase transitions between superconducting states of potentially different pairing symmetry. Furthermore, it might be expected that our observation is not limited only to oxide interfaces but can be extended to other material platforms with flat bands and spin-orbit coupling.

\textbf{Acknowledgment} We thank Y. Yanase and N. Nagaosa  for fruitful discussions. A.E. acknowledges financial support from DFG through priority program SPP1666 (Topological Insulators), SFB-TRR227,  and OeAD Grants No. HR 07/2018 and No. PL 03/2018.  E.Ya.S. acknowledges financial support through the Grant PGC2018-101355-B-I00 funded by MCIN/AEI/10.13039/501100011033 and by ERDF ``A way of making Europe'', and by the Basque Government through Grant No. IT986-16. M.Kawasaki acknowledges financial support from JSPS KAKENHI 22H04958.

\textbf{Author Contributions:} 
D.M. conceived the project, grew structures and performed experiments with the help of M. Kawamura.  M. Kriener, D. Z. and M. Kawamura contributed to the discussion in the course of experiments. All authors discussed the results. I.V.M., S.O. and A.E. performed band structure calculation. V.K.D and E.Ya.S. developed the theoretical model. M. Kawasaki supervised the project. D.M. wrote the manuscript with contribution from  V.K.D., E.Ya.S, I.V.M. and S.O.  All authors commented on the manuscript. 
%

\end{document}